# Structural, elastic, optoelectronic and transport properties of Sr₃SnO under pressure

Enamul Haque and M. Anwar Hossain*

Department of Physics, Mawlana Bhashani Science and Technology University, Santosh, Tangail-1902, Bangladesh

Email: anwar647@mbstu.ac.bd, enamul.phy15@yahoo.com

**Abstract**

In this paper, we have presented the structural, elastic, optoelectronic and transport properties of $Sr_3SnO$ (SSO) under pressure by using first-principles method. The application of hydrostatic pressure causes charge transfer from Sr(5s) orbital to Sn(5p) and O(2p) orbitals. The increasing trend of Pugh's ratio (B/G) under pressure implies that the material tends to be ductile at high pressure. The semiconductor-metal transition occurs at 14 GPa and the density of states at the Fermi level is significantly increased at this pressure. The refractive index, optical conductivity, and absorption of SSO have been found to be high and comparable to that for typical materials used in photovoltaic. The material becomes n-type from 12 GPa and Hall coefficient also confirm it. The Seebeck coefficient is still high (-111.73 µV/K at 12 GPa and 360 K). Thus, SSO is a potential thermoelectric material possessing both p- and n-type nature. The detail physics of these changes under pressure has been explained within the available theory.

## 1. Introduction

The antiperovskite Dirac metal oxides (ADMOs), in which metal ions have negative valence states [1], exhibit some unusual physical properties such as superconductivity [2–4], topological behavior [5], ferromagnetic behavior [6–10]  and thermoelectric transport properties [11,12]. In ADMOs, the metal ions reverse their positions to the oxygen ions [12]. The $Sr_3SnO$ (SSO) (an ADMO) is a nearly topological insulator that has been recently found to exhibit topological superconductivity due to Sr deficiency, with a superconducting transition temperature ($T_c$) ~5 K [2,4,5]. The strong hybridyzation between Sr-4d and Sn-5p orbitals has been found to be responsible for topological superconductivity in SSO [2,3]. However, many theoretical and experimental studies reported that SSO show nearly topological insulating and semiconducting behavior [4–6]. Recently, the SSO with Si (001) has found to be dilute magnetic semiconductor [6,8]. The theoretical studies on $A_3SnO$ ($A$ = Ca, Sr, Ba)  have been found that these  ADMOs possess high thermopower [12]. Furterore,  $Ca_3SnO$ has been experimentally found to possesses high thermopower, which approaches about ~ 100 µV/ K at the room temperature [11]. For practical applications, the mechanical stability against external pressure is the important criterion. Recently, it has been found that thermoelectric performance of BiCuSeO, $Bi_2Te_3$, PbTe and can be improved by the external pressure [13–15]. Furthermore, the study of optical properties of a material is essential to reveal its suitability in the photovoltaic applications [16–18].   These facts give rise interest to study the eleastic, electronic, optical and transport properties of SSO under hydrostatic pressure. The pressure can change the molcular orientation and also the phase of the material. Furthermore, the superconductivity can be suppressed or enhanced by the pressure [19–21]. The electronic structure can significantly change due to the applied pressure. For example, the bandgap can increase [13,22] or decrease [23] by the external pressure. The SSO crystallize in cubic structure with lattice paramter, a=5.1394 Å and space group $Pm\bar{3}m$  (#221) [1]. The Wyckoff positions for Sr, Sn and O atoms are 3c (0, 0.5, 0.5), 1a (0, 0 ,0), and 1b (0.5, 0.5, 0.5), respectively [1,24].

In this paper, we present the elastic, optoelectronic, and transport properties of $Sr_3SnO$ under pressure by using the first-principles method within density functional theory [25,26]. We have found that SSO becomes metals above 12 GPa pressure (at low temperature, below 50 K) and n-type material. Above this pressure, the Seebeck coefficient (S) still remains high although metallic conductivity observed at low temperature. Since thermoelectric generator requires both p-type and n-type materials [27–29], therefore, SSO is a promising material for thermoelectric generator applications. The refractive index of SSO has been found to be 3.14\ which is smaller than that the value (5.974) for Ge [30] but close to the value for GaAs (3.29-3.857) [31,32] at 632 nm wavelenth and 0 GPa pressure.

## 2.      Computational method

The geometry optimization by under pressure was studied by the plane wave pseudopotential method in CASTEP [33]. The Mulliken population [34] was also studied in CASTEP. In these calculations, the generalized Perdew-Burke-Ernzerhof (PBE-GGA) functional [35,36], 21 × 21 × 21 k-point for BZ integration, and 500 eV cutoff energy were used. The elastic constants and related parameters were calculated by using IRelast method [37] interfaced with WIEN2k. The electronic structure and optical properties were investigated by using a full potential augmented plane wave method (FP-LAPW) as implemented in WIEN2k [38]. The results obtained by two different code were crosschecked while it was possible. Both codes were found to produce identical results. To overcome the underestimation of bandgap (PBE-GGA [39]), the electronic structure was calculated by using Tran-Blaha modified Becke-Johnson potential (TB-mBJ) [40,41]. The BZ integration was performed within 21 × 21 × 21 k-point. The muffin tin radii: 2.3, 2.5, 2.3, and 2.2 Bohr for Sr, Sn, O and Ca, respectively and the kinetic energy cutoff ($RK_{max}$) 7.0 were chosed for good convergence of basis set. The core states and valence states seperation energy was set to -8.0 Ry due to deep lying Sn-5p states. The energy and charge convergence criteria were chosed to be $10^{-4}$ Ry and 0.001e. Since both optical and transport properties calculations require denser k-point, the self-consistent field (SCF) calculation was performed again with $47 \times 47 \times 47$ k-point. The transport properties were studied by solving Boltzmann transport equation as implemented in BoltzTraP [42]. The values of transport parameters were taken at the value of chemical potential equal to 0 K Fermi energy.

## 3.    Result and Discussions

### 3.1.    Structural properties

The cubic SSO structure contains three nonequivalent atoms per unit cell as shown in the Fig. 1. Thus, internal atomic parameter Z=1. The number atomic bond in SSO is 3. The first one is between O and $Sr_1$, the second one is $O-Sr_2$ and the third is $O-Sr_3$.

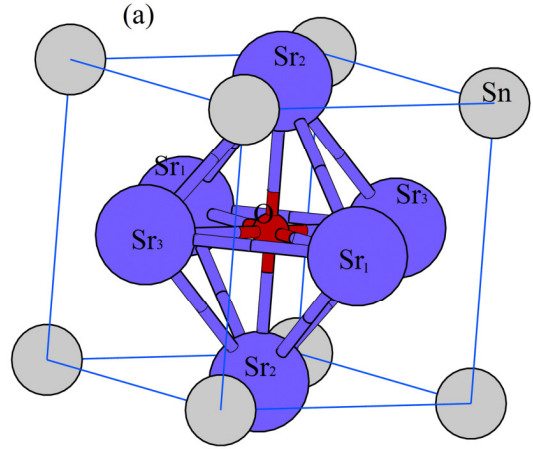

Fig. 1: The crystal structure of $Sr_3SnO$

All of these bonds have the same length and the value is 2.586 Å. The optimized lattice parameter at 0 GPa has been found to be 5.1473 Å, with excellent agreement with the experimental value 5.1394 Å [1]. The lattice parameter decreases linearly with the increasing pressure as illustrated in Fig. 2(a). Since the pressure reduces the interatomic distance, the bond length also decreases linearly with it (see Fig. 2(b)).

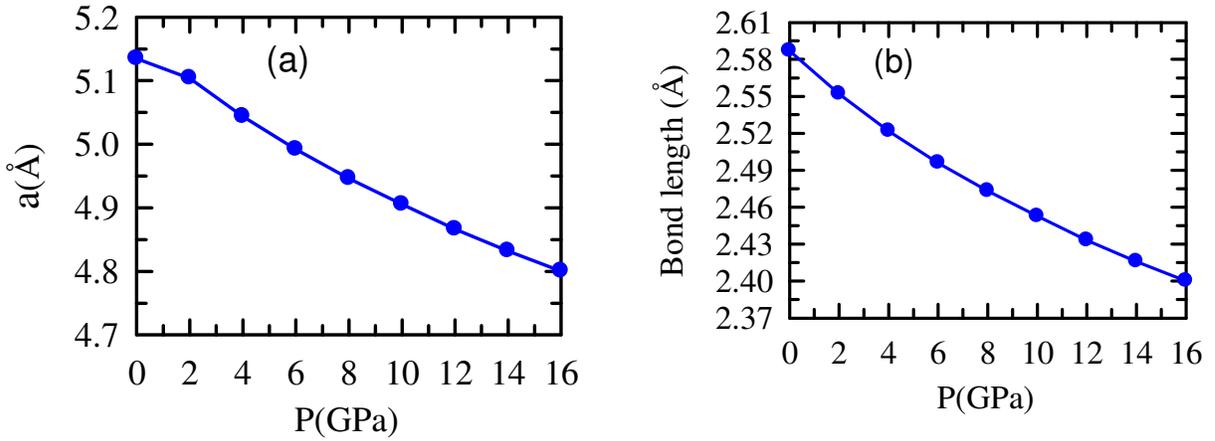

Fig. 2: Variations of structural parameters with pressure: (a) lattice parameters, and (b) bond length.

The pressure also influences the atomic orbits and hence the charge density. The charge density of SSO at the ambient pressure is illustrated in the Fig. 3(a). The light red circles inside the light green circle indicate the positive charge of Sr, while only the light green circles (smaller than Sr circle) indicate the negative charge of O, as clarified by the corresponding symbol. The four point star symbol present the negative charge of Sn. Some charges of Sr transfer to Sn and O due to the increase of pressure as clearly indicated in the Fig. 3(b). The partially filled 5p orbitals of Sn and 2p orbitals of O tend to get more electrons to be more stable. The 5s orbitals of Sr tend to leave its outer shell electrons to be in its equilibrium configuration. Thus, the valence charge of Sr transfer to Sn and O with the increase of pressure.

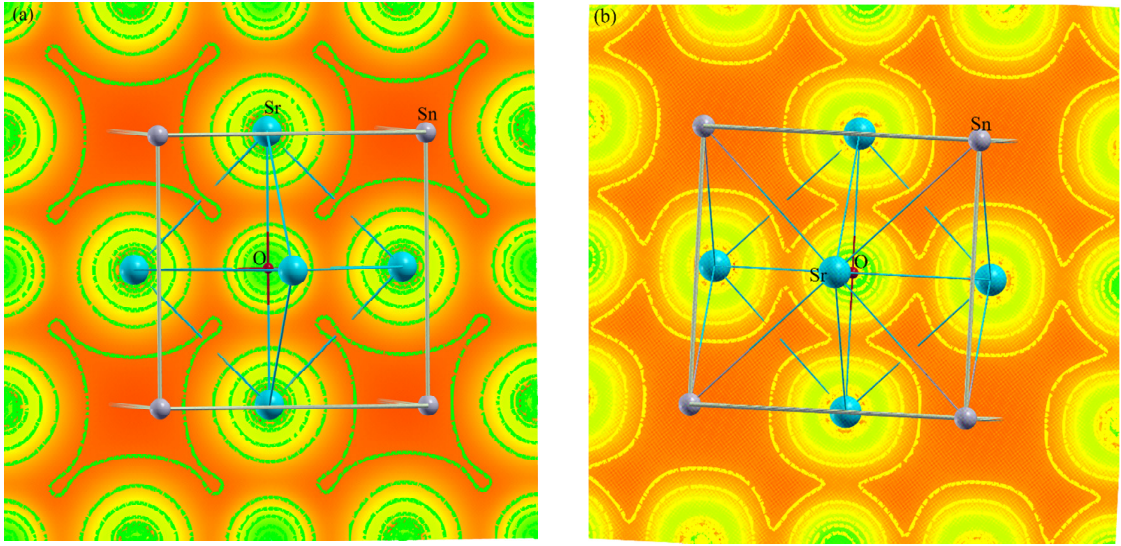

Fig. 3: The charge density (e/Å³) mapping at: (a) 0 GPa and (b) 14 GPa pressure. These figures are sketched in the (20 1 1) plane with Xcrysden program [43].

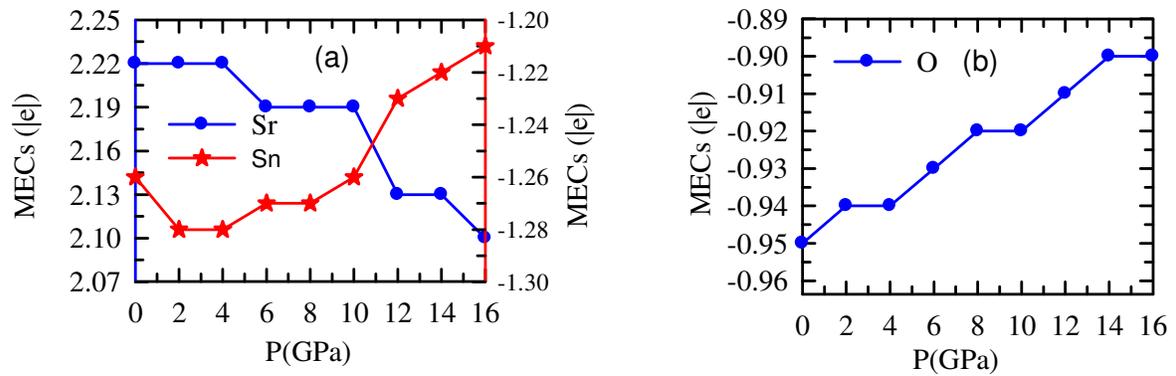

Fig. 4: Variations of Mulliken effective charges (MECs) with pressure: (a) for Sr, Sn, and (b) O.

For further clarification, we have calculated the atomic charges to the corresponding pressure from Mulliken population analysis. The variation of Mulliken effective charges of Sr, Sn and O atoms under pressure are presented in Fig. 4. The positive Mulliken effective charges (MECs) of Sr decrease with the increase of pressure while the negative MECs of Sn and O increases. This indicates the pressure induced charge transfer from the Sr-5s orbitals to Sn-5p and O-2p orbitals. Such charge transfer can induce soft phonon mode [44], as this can change the activity between attractive and repulsive forces. Furthermore, the attractive forces are weakened due to the charge transfer with increasing pressure. The most important fact that such phonon-softening mode due to the pressure induced charge transfer may enhance the electron-phonon coupling strength in SSO. Further study is required to clarify this predication.

### 3.2. Elastic properties

Since the bond length decreases linearly with increasing pressure, it is interesting to study the effects of pressure on elastic properties. The three elastic constants, $c_{11}$, $c_{12}$, and $c_{14}$ are used to describe the elastic properties of a cubic system. The first two elastic constants increases linearly and rapidly, while the third constant increase slowly, as shown in the Fig. 5(a). The bulk modulus B can be calculated from elastic constants as given by $B = (c_{11} + c_{12})/3$. Since the bulk modulus is related to the resistance to the external effects, the rapid increase of $c_{11}$ and $c_{12}$ with pressure indicates the increase of resistance to the hydrostatic pressure. However, $c_{14}$ is related to the shear modulus (G) defined as $G = 1/2[(c_{11} - c_{12} + 3c_{44})/5 + 5c_{44}(c_{11} - c_{12})/(4c_{44} + 3(c_{11} - c_{12}))]$. Since the material surface changes slowly with the applied hydrostatic pressure, $c_{14}$, thus, also increases slowly. The elastic stability conditions, under pressure P, can be generalized as follows [45,46]

$$\left[\frac{1}{2}(c_{11} - c_{12} - 2P)\right] > 0$$

$$\left[\frac{1}{3}(c_{11} + c_{12} + P)\right] > 0 \ and \ [(c_{44} - P)] > 0.$$

Our calculated elastic constants within the pressure range 0-16 GPa indicate that SSO is elastically stable under considered pressure. By using bulk and shear modulus, the Young modulus and Poisson ratio can be calculated by using the standard expressions [47–49]. The values of these parameters at different pressure are presented in the Table-1. We can get an idea about the average bond strength from the bulk modulus [50].

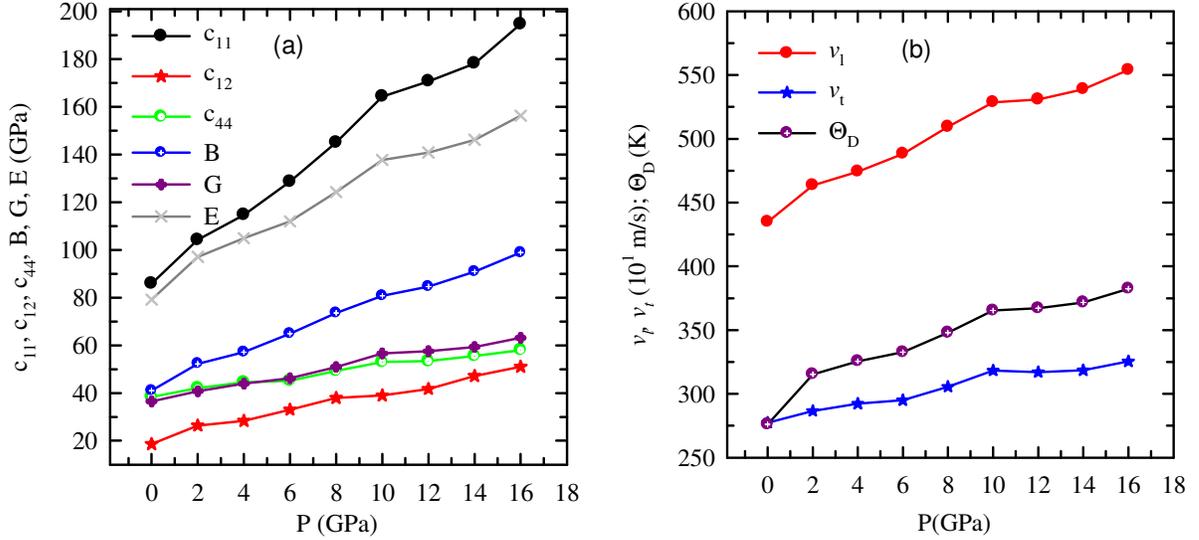

Fig. 5: The change of (a) elastic constants, moduli of elasticity and (b) longitudinal (vl), transverse (vt) sound velocity, and Debye temperature at a different pressure.

Table-1: The calculated elastic parameters (GPa), Cauchy pressure ($c_{12} - c_{44}$ in GPa), Pugh ratio (B/G), Poisson ratio ($v$), and predicted Vickers hardness (H$_V$) at different pressure.

| P | $c_{11}$ | $c_{12}$ | $c_{44}$ | $c_{12} - c_{44}$ | B | G | $v$ | B/G | A | H$_V$ |
|---|---|---|---|---|---|---|---|---|---|---|
| 0 | 85.90 | 18.60 | 38.40 | -19.80 | 41.00 | 36.50 | 0.160 | 1.12 | 1.14 | 12.33 |
| 2 | 104.25 | 26.37 | 42.03 | -15.66 | 52.33 | 40.76 | 0.190 | 1.28 | 1.08 | 12.12 |
| 4 | 114.69 | 28.37 | 44.52 | -16.15 | 57.15 | 43.97 | 0.193 | 1.30 | 1.03 | 12.69 |
| 6 | 128.63 | 33.07 | 45.16 | -12.09 | 64.92 | 46.21 | 0.212 | 1.41 | 0.95 | 12.43 |
| 8 | 145.06 | 37.96 | 49.31 | -11.35 | 73.66 | 50.96 | 0.218 | 1.45 | 0.92 | 13.08 |
| 10 | 164.33 | 39.00 | 52.98 | -13.98 | 80.78 | 56.66 | 0.216 | 1.43 | 0.85 | 14.24 |
| 12 | 170.72 | 41.74 | 53.38 | -11.64 | 84.74 | 57.58 | 0.222 | 1.47 | 0.83 | 14.09 |
| 14 | 178.26 | 47.18 | 55.57 | -8.39 | 90.87 | 59.36 | 0.232 | 1.53 | 0.85 | 14.00 |
| 16 | 194.64 | 51.04 | 58.02 | -6.97 | 98.91 | 63.18 | 0.236 | 1.56 | 0.81 | 14.39 |

The increase of bulk modulus with increasing pressure (See Fig. 5(a)) indicates that the bond strength increases, while bond length decreases. Therefore, the resistance to the volume deformation under hydrostatic pressure increases. The shear modulus (G) determines how much the rigidity of a material is. The slow increase of G with the pressure indicates that the rigidity of SSO also increases slowly. It is difficult to relate the hardness to the shear modulus directly [51] and the Pugh's ratio indicates that both bulk and shear moduli are essential to describe the hardness of a material [52]. However, the hardness can be calculated by Chen's formula [53] $H_V = 2 \left( \left( \frac{G}{B} \right)^2 G \right)^{0.585} - 3$. It is clear from the Table-1 that hardness increases with pressure as expected. The Pugh's ratio increases with the increase of pressure indicating that the material tends to be ductile at high pressure. The negative Cauchy pressure ($c_{12} - c_{44}$) [54] decreases with the increase of pressure giving further confirmation of brittle to ductile transition at high pressure. The negative Cauchy pressure also indicates the presence of directional covalent bonds in SSO. The calculated Poisson ratio (ductile –brittle phase is separated by the critical value $v \sim 0.26$) at different pressure gives the same confirmation [55]. The Poisson's ratio measure the volume change during uniaxial elastic deformation. The larger values implies the smaller change in volume and vice versa. The elastic deformation cannot bring any change in volume of a material possessing Poisson ratio 0.5. The small value of Poisson ratio indicates larger volume of SSO change will be caused by the elastic deformation. Poisson ratio can also describe the nature of bonding [56]. The upper and lower limit of Poisson ratio for central force are 0.25 and 0.5, respectively. The upper limit implies the infinite anisotropy of the material [57]. The Poisson ratio of SSO indicates that the interatomic forces are weakly central with finite anisotropy. The anisotropic nature of a material can be further described by the relation [58] $A = \frac{2c_{44}}{c_{11} - c_{12}}$. From the Table-1, we see that the shear anisotropy decrease with the increase of pressure implying that SSO tends to be elastically isotropic solid.

The Debye temperature is an important parameter related to thermal stability, lattice thermal conductivity, superconductivity (appeared in the McMillan formula [59]), etc.. The method of calculation of Debye temperature can be found in the standard article [60]. From the Fig. 5(b), we see that the longitudinal and transverse sound velocity through SSO increases with pressure. Since the atoms in a crystal can be considered as the chain and pressure reduces the interatomic distance, the propagation velocity of vibrational sound waves increases as the atoms become closer with the increase of pressure. The pressure changes the length more than the surface, thus, longitudinal sound velocity increases more rapidly than the transverse wave. Since the Debye temperature is directly proportional to the average sound velocity through the solid, the Debye temperature increases with the increase of pressure.

### 3.3.    Electronic properties

Since the bond length decreases with pressure, as seen in the previous section, the electronic structure of SSO should change significantly. The pressure should stabilize the bonding orbitals while the antibonding orbitals should be destabilized. The pressure may lead to the semiconductor to metal transition (as found for Si [61]), superconductivity, etc.. In order to find the pressure effect on the electronic structure we have calculated the band structure and density of states of SSO. The band structure of SSO at different pressure is shown in the Fig. 6. We see that the basic features of the band structure of SSO remain unchanged with pressure. The pressure shifts the Fermi level toward the conduction band. Thus, we see that the pressure has a similar effect like doping [62]. In previous study, we have found that the Fermi level shifts to higher energy due to Ca substitution [62]. Above 12 GPa pressure, the maxima of the valence bands (MVBs) and minima of the conduction bands (MCBs) overlap at the $\Gamma$-point as shown in Fig. 6(d).

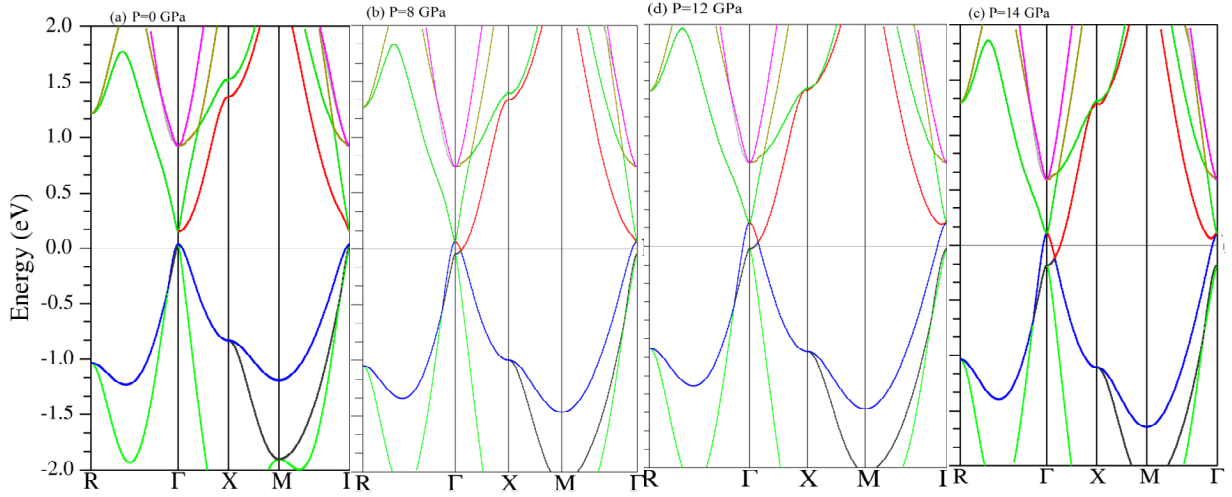

Fig. 6: The calculated band structure of SSO by using TB-mBJ functional at different pressure: (a) 0 GPa, (b) 8GPa, (c) 12 GPa, and (d) 14 GPa. The straight line at the zero energy represents the Fermi level.

This is may be due to the stabilization of bonding orbitals with pressure. At 8 and 12 GPa pressure, a very small gap exists between MVBs and MCBs at Γ-point. Such semiconductor to metal transition due to the pressure can significantly change other physical properties related to the electronic structure of SSO, such as optical and transport properties.

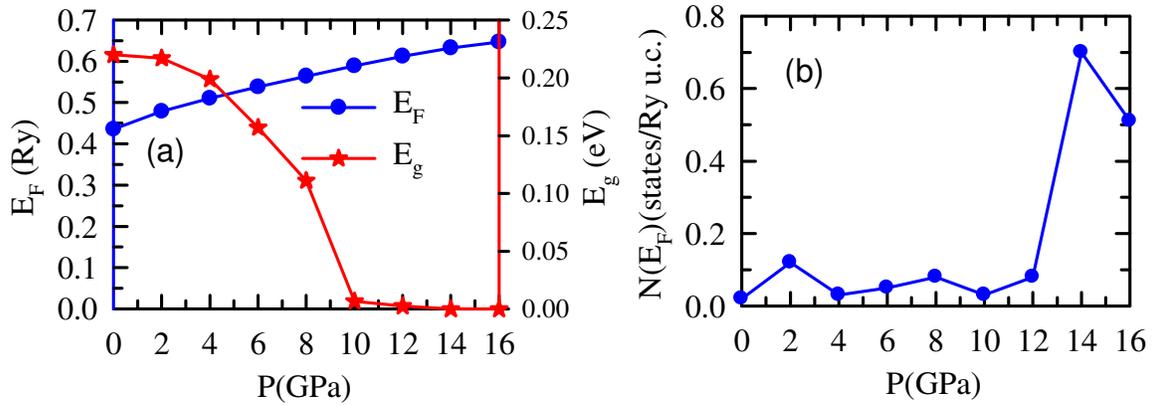

Fig. 7: Variations of basic electronic structure parameters with pressure: (a) Fermi energy ($E_F$), (b) bandgap ($E_g$), and density of states at the Fermi level.

For further clarification of the effects of pressure on the electronic structure of SSO, we have illustrated the variations of Fermi energy, bandgap, and density of states at the Fermi level in Fig.7. We see that Fermi energy linearly increases with pressure. The Fermi energy is related to carrier density. Since the pressure effect transfer charges from Sr (5s) to Sn(5p) and O(2p) orbitals, the carrier density increases and hence Fermi energy increases. The bandgap decreases with pressure and a sharp drop occurs at 10 GPa as shown in Fig. 7(a) (with a red line and star symbol). The bandgap becomes zero at 14 GPa at which the density of states at the Fermi level is maximum as shown in Fig. 7(b). The calculated total and atomic density of states of SSO at different pressure are shown in Fig. 8. The basic shape of the total density of states remains unchanged with pressure (see Fig. 8 (a)).

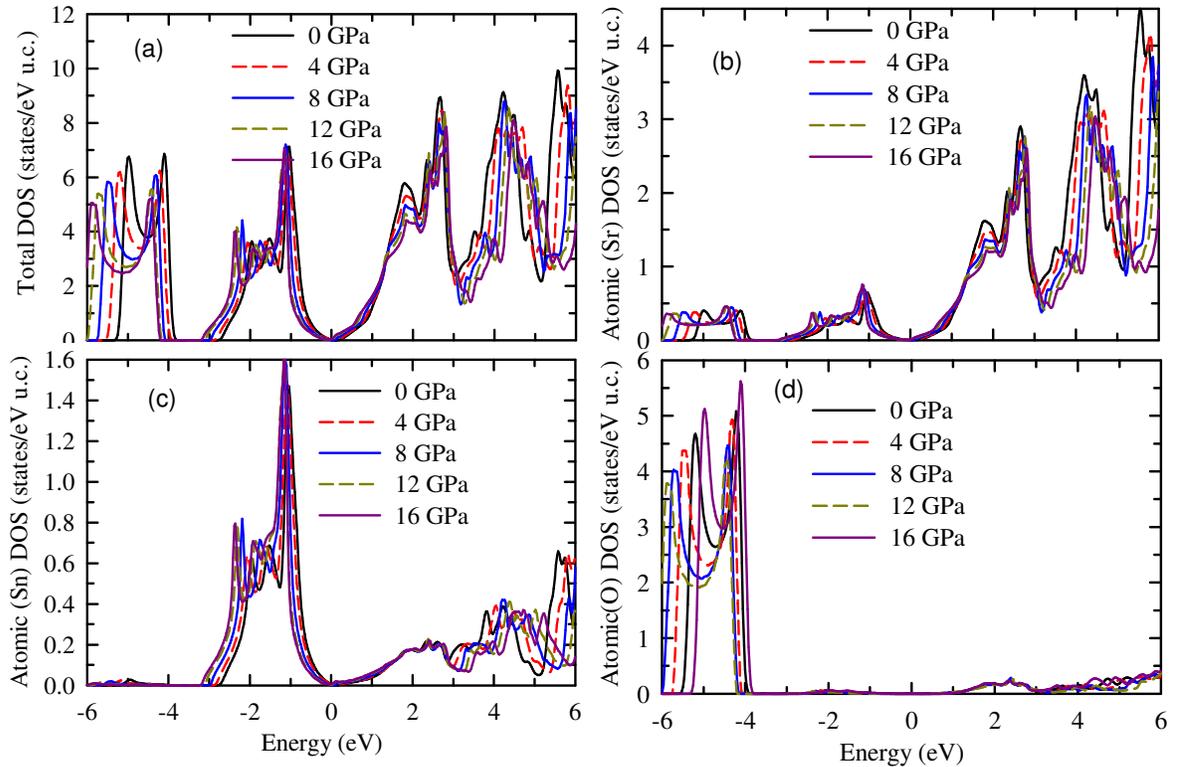

Fig. 8: The calculated total and atomic density of states (DOS) of SSO: (a) total DOS of SSO, (b) DOS of Sr, (c) DOS of Sn, and (d) DOS of O.

We see that the density of states of valence bands decrease with pressure while the DOS of conduction bands increases around the Fermi level. The DOS of Sr decreases with the increase of pressure as illustrated in the Fig. 8(b). However, the DOS of the conduction bands of Sn and O around Fermi level increases. Therefore, the bonding orbitals get more stabilized with pressure while the antibonding orbitals become less stable. These indicate that due to pressure effect Sr becomes more stable getting full outer electrons shells. But Sn and O share the small charges of Sr ($5s^2$) almost equally and hence they cannot gain full outer electrons shells to be in a stable configuration, rather becomes more unstable. Thus, the density of states of conduction bands around Fermi level increases with the increase of pressure. Since the main contributions to the DOS of SSO come from the strong hybridization between Sr-4d and Sn-5p orbitals [2,3], the hybridization gets stronger with the increase in pressure.

### 3.4.    Optical properties

The change of electronic structure due to pressure should produce a significant change in the optical response of SSO. The optical response of a medium for photon energies E=ℏω can be described in terms of frequency dependent complex dielectric functions $\varepsilon(\omega) = \varepsilon_1(\omega) + i\varepsilon_2(\omega)$. The imaginary part of the dielectric function can be obtained from the electronic structure calculation, within the Ehrenreich and Cohen approach [63]. In this approach, the imaginary part of the dielectric function is defined as

$$\varepsilon_2(\omega) = \frac{e^2 h}{2\pi^2 m^2 \omega^2} \sum_{v,c} \int |M_{cv}(k)|^2 \, \delta[\omega_{cv}(k) - \omega] d^3 k$$

where $M_{cv}$ is the transition momentum matrix and the integration is performed over the first BZ. Then the real part can be obtained by using Kramers–Kronig relation [64,65]

$$\varepsilon_1(\omega) = 1 + \frac{2}{\pi} P \int_0^\infty \frac{\omega' \varepsilon_2(\omega')}{\omega'^2 - \omega^2} d\omega'$$

where P is the Cauchy principal value of improper integral. Then, we can easily calculate the important optical parameter, refractive index n(ω),

$$n^2(\omega) = \frac{1}{2} \left[ \varepsilon_1(\omega) + \sqrt{\varepsilon_1^2(\omega) + \varepsilon_2^2(\omega)} \right]$$

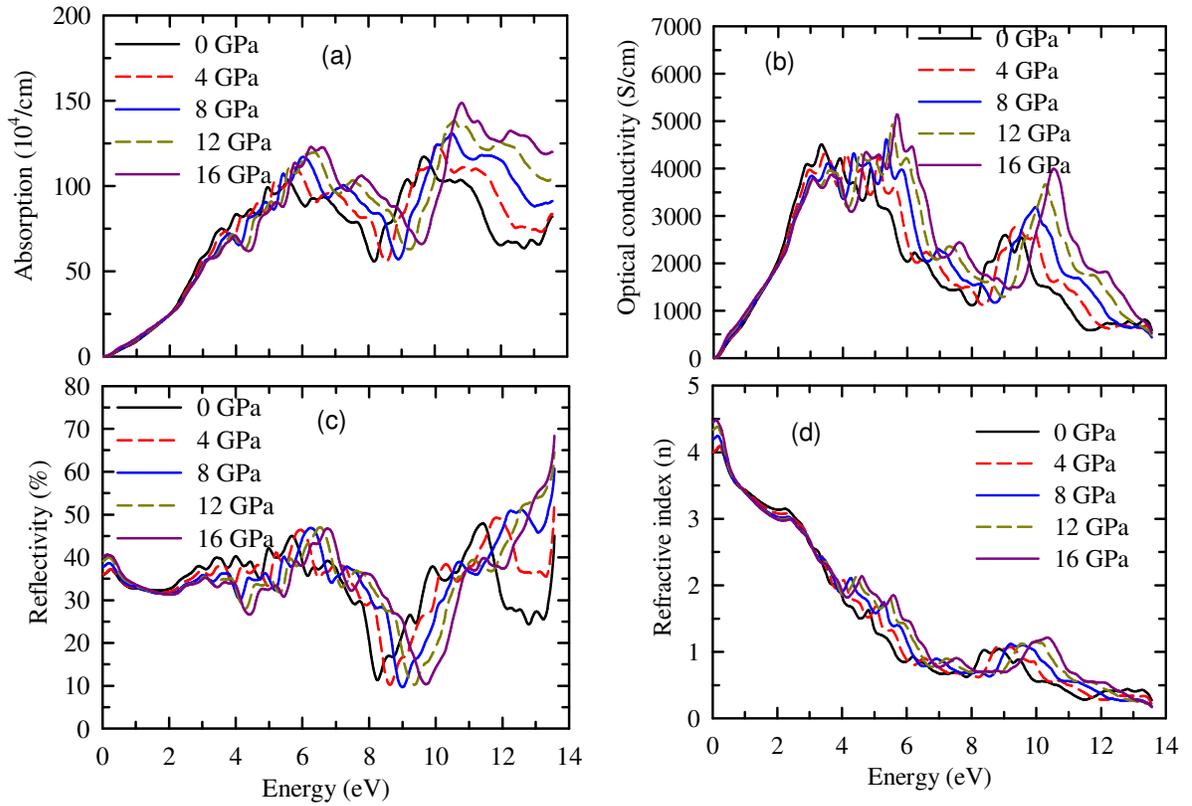

Fig. 9: The variations of (a) absorption, (b) optical conductivity, (c) reflectivity, (d) refractive index of SSO with phonon excitation energy (eV) at different pressure. No intraband contribution was added since intraband contribution for semiconductor is negligibly small [66–68].

The details of the calculation of all optical parameters have been described by Ambrosch- Draxl *et. al.* [69]. The material with high absorption, high optical conductivity, low emissivity, and high refracive index, can be used in photovotaic systems [18,70]. The calculated absorption, optical conductivity, relectivity, and refractive index of SSO at a different pressure within the photon excitation energy 0-14 eV are shown in Fig. 9. The maximum peak of absorption is obatined at 11 eV photon energy and 16 GPa. We see that the absorption increases with pressure for certain energy range and also decreases for other energy. This is true for other optical parameters of SSO. We see that the optical conductivity of SSO is high and the maximum value is 4465.5 S/cm at 6.13 eV and 16 GPa. The reflectivity of SSO is small compairatively to other typical semiconducotors, Si, Ge, and GaAs [71]. The refrative index of SSO at different pressure is illustrated in the Fig. 9 (d). We see that the refractive index of SSO is smaller (3.139) than that of Ge (5.974) [30,72–74] but very close the value of GaAs (3.29-3.857) [31,32] (at 632 nm wavelength and 0 GPa).

The joint density of states (JDOS) is defined as [75] $J(\omega) = \frac{1}{2\pi^2} \left( \frac{2\mu_m}{\hbar} \right)^{3/2} \left( \omega - \omega_g \right)^{1/2}$, where $\mu_m$ is the reduced effective mass and the bandgap energy is given as E=$\hbar\omega_g$. The JDOS determines the possible number of pairs states separated by the energy ℏω. These states must have the same wave vector. The calculated Joint density of states is illustrated in Fig. 10(a). The peak at the 3 eV photon energy arises from the conduction bands and valence bands splitting at the Γ-point. The imaginary part of the dielectric function is shown in Fig. 10 (b). The real part of the dielctric function has also been calculated but not presented here. The maximum peak of dielectric function is at 3 eV which is consistent with calculated JDOS. This peak gives the threshold for the direct optical transition between VBM and CBM. This energy is refered to as the fundamental absorption edge. Fig. 10 (c) shows the variations of energy loss function with photon energy at different a pressure. The energy loss function is relatively smaller than ZnSe and LiMn$_2$O$_4$ [76,77]. The small energy losss function implies that few number of electrons will undergo inelastic scattering.

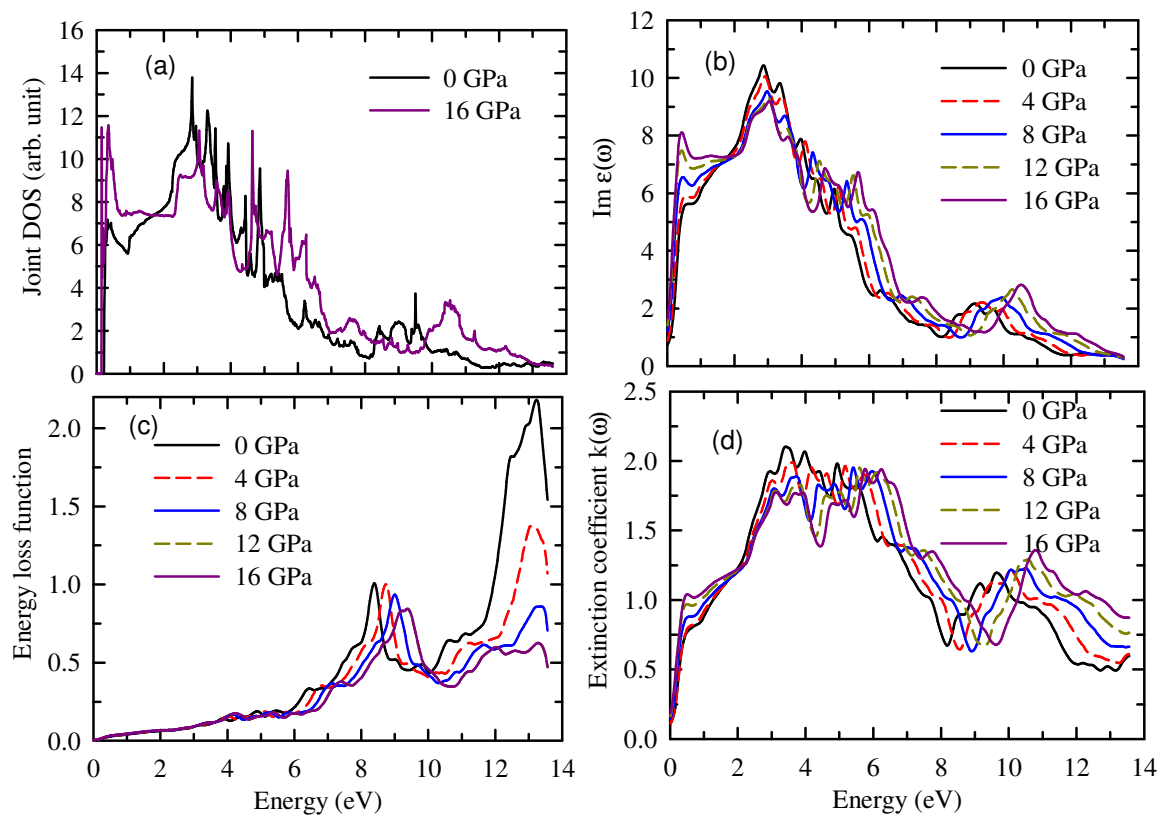

Fig. 10: The changes of (a) Joint DOS, (b) imaginary part of the dielectric function, (c) energy loss function, and (d) extinction coefficient, with photon energy at different pressure.

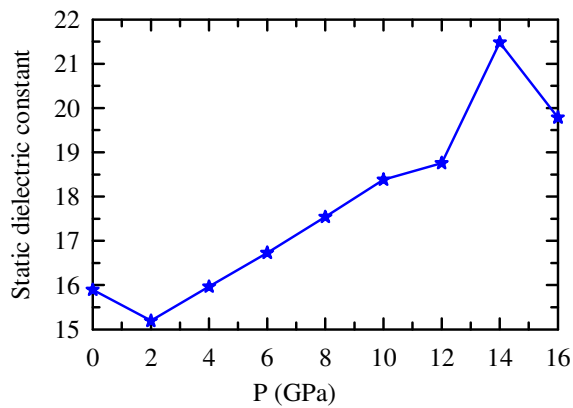

Fig. 11: The variations of static dielectric functions with pressure

The absorption coefficient of a material is directly proportional to the extinction coefficient, the imaginary part of the complex refractive index [78]. The large value of the extinction coefficient indicates high absorptivity of SSO. The variations of static dielectric constant (SDC) with pressure is shown in the Fig. 11. We see that the SDC decreases from 0 to 2 GPa, and then increases linearly. At 14 GPa, the sharp increase of SDC occurs and this implies the semiconductor to metal transition, which is consistent with our electronic structure calculation. The above analysis of the optical properties of SSO reflects the fact that SSO is a potential candidate for photovoltaic applications. Further experimental study on the optical properties of pristine SSO is needed for further confirmation of our obtained results.

### 3.5.    Transport properties

The energy bandgap of SSO is reduced with pressure and the carrier density of conduction bands increases as seen in the previous section. Therefore, the thermoelectric transport properties should be drastically changed with pressure. The Seebeck coefficient ($S$) determines how much emf will be generated across the junctions when there exist temperature gradient and the carrier types. The calculated Seebeck coefficient (within constant relaxation time approximation) of SSO is shown in the Fig. 12(a). The Seebeck coefficient increases with increasing temperature until the intrinsic activation energy of the material approaches to the Fermi level. The $S$ at 0 GPa increases linearly with temperature upto 420 K and then decreases gradually, i. e., activation energy becomes equal to the Fermi level at 420 K, for which the Fermi level is shifted to the middle of the bandgap. When it is the pressure 4 GPa, the $S$ increase linearly over the whole considered temperature range. Since Fermi energy increases with pressure, thus, much high activation energy requires approaching the Fermi level, which in turns high temperature is required for this.

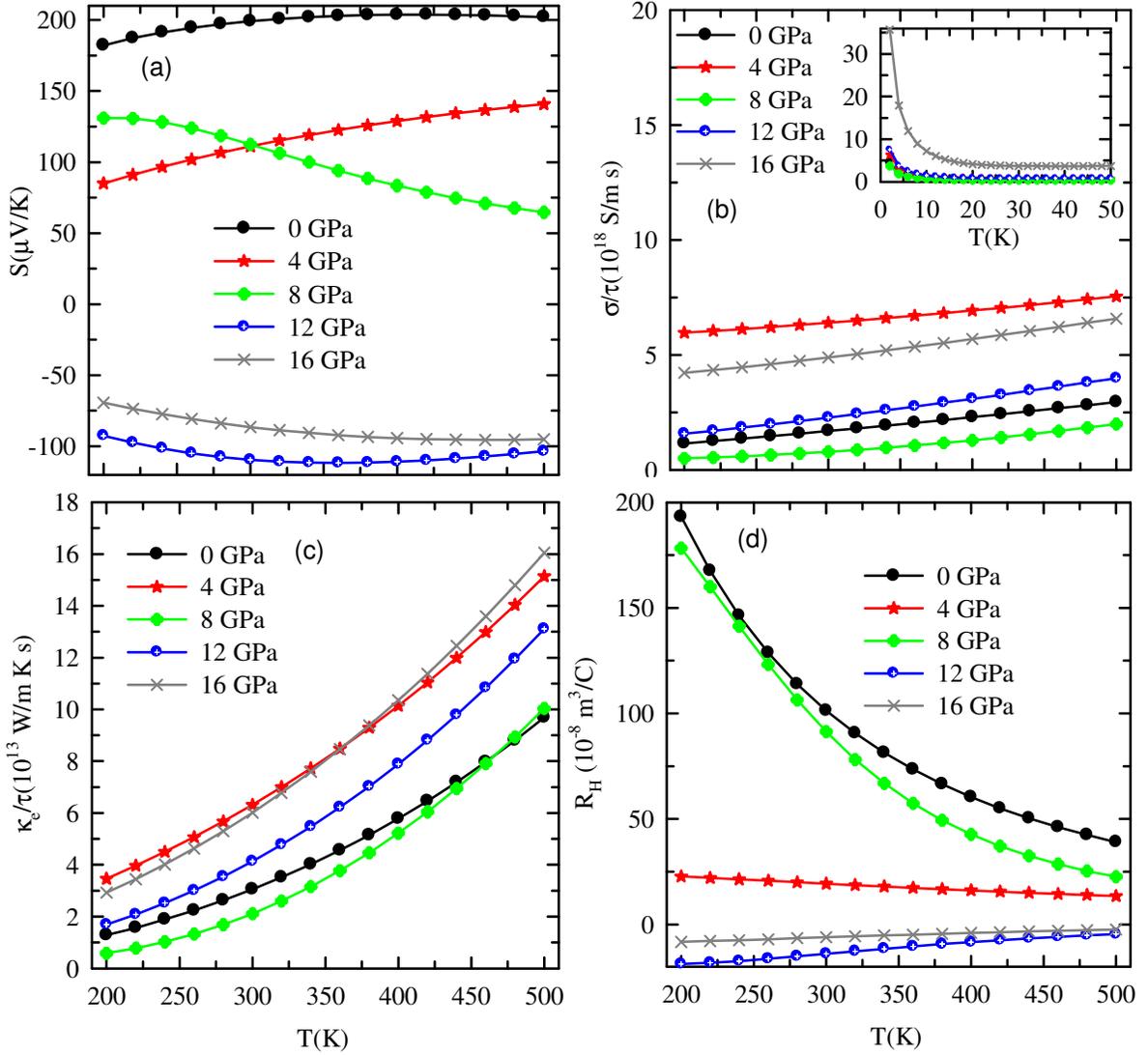

Fig. 12: The variations of transport parameters with the temperature at different pressure: (a) Seebeck coefficient ($S$), (b) electrical conductivity ($\sigma/\tau$), (c) electronic part of the thermal conductivity ($\kappa/\tau$), (d) Hall coefficient $R_H$.

Surprisingly, when we increase the pressure to 8 GPa, the Seebeck coefficient start to decrease at low temperature (from 200 K). This is because the carrier density around the Fermi level significantly increase and hence the activation energy increases. Therefore, the activation energy easily approaches the Fermi level at low temperature. When we further increase the pressure, the Seebeck coefficient becomes negative and increases slowly. Above 12 GPa the bandgap becomes zero and the density of states significantly increases as seen from electronic structure section. This implies that the electrons density increases while holes density decreases, therefore, the SSO becomes n-type material above 10 GPa. This is a great advantage of SSO to use in the thermoelectric generator (TEG) since the TEG module requires both n- and p-type material. Although the bandgap nearly zero at 12 GPa, the Seebeck coefficient still remains high, approaching to the maximum value -111.73 μV/K at 360 K. We see that the Hall coefficient also changes its sign at 12 GPa, which is consistent with the carrier density and electronic structure calculation. Surprisingly, the electrical conductivity increases with the increasing temperature at all the considered pressure range and this implies the semiconducting nature, as shown in the Fig. 12 (b). However, at low temperature, SSO shows metallic conductivity, for all considered pressure, from 2K to 50K (32 K for P=0GPa). Since phonon scattering is small at low temperature and thus the relaxation time is large, therefore, the constant relaxation time approximation cannot give a correct trend of the conductivity at low temperature [79–81]. Since the electrical conductivity increase with temperature and the Seebeck coefficient is high at 12 GPa, the power factor must be high, indicating suitability of SSO in TEG module applications. The electronic thermal conductivity of SSO increases with increases temperature as usually, due to the increase of carrier density, as shown in the Fig. 12(c). The electronic part of the thermal conductivity also increases with pressure as the pressure causes the carrier density to be increased.

## 4. Conclusions

In summary, the effect of pressure on the structural, elastic, optoelectronic and transport properties of $Sr_3SnO$ (SSO) has been studied by using first-principles method. The lattice parameter and bond length of SSO decrease linearly with pressure. The pressure effect causes the charge transfer from Sr (5s) orbital to Sn (5p) and O (2p) orbitals. The elastic moduli and Debye temperature have been found to increase with pressure as it reduces the interatomic distance. The Pugh's ratio (B/G) increases with pressure and implies that the material tends to be ductile at high pressure. The bandgap decreases while Fermi energy increases with pressure. The semiconductor to metal transition occurs at 14 GPa and the density of states at the Fermi level is significantly increased at this pressure. The refractive index of SSO is 3.139, which is smaller than that of Ge (5.974) but close to the value of GaAs (3.29-3.85), at 632 nm wavelength and 0 GPa pressure. The optical conductivity and absorption of SSO have been found to be high and comparable to that for typical materials used in photovoltaic. The Seebeck coefficient decreases with pressure and becomes negative from 12 GPa. Thus, SSO become n-type material from 12 GPa and negative Hall coefficient also confirm it. The Seebeck coefficient is still high (-111.73 µV/K at 12 GPa and 360 K). Thus, SSO is a potential thermoelectric material possessing both p- and n-type nature. The physical reasons behind these changes have been explained in details.